# SCIENTIFIC REPORTS

natureresearch

**OPEN**

# Tuning the Néel temperature in an antiferromagnet: the case of $Ni_xCo_{1-x}O$ microstructures

Anna Mandziak[1,2], Guiomar D. Soria[1], José Emilio Prieto[1], Pilar Prieto[3], Cecilia Granados-Miralles[4], Adrian Quesada[4], Michael Foerster[2], Lucia Aballe[2] & Juan de la Figuera[1]

We show that it is possible to tune the Néel temperature of nickel(II)-cobalt(II) oxide films by changing the Ni to Co ratio. We grow single crystalline micrometric triangular islands with tens of nanometers thickness on a Ru(0001) substrate using high temperature oxygen-assisted molecular beam epitaxy. Composition is controlled by adjusting the deposition rates of Co and Ni. The morphology, shape, crystal structure and composition are determined by low-energy electron microscopy and diffraction, and synchrotron-based x-ray absorption spectromicroscopy. The antiferromagnetic order is observed by x-ray magnetic linear dichroism. Antiferromagnetic domains up to micrometer width are observed.

Antiferromagnetic materials (AFM) have a great potential in the next generation of spintronic devices[1]. Already in current devices such as magnetic tunnel junctions and high-density memories, antiferromagnetic materials, sometimes of nanometer thickness[2], are used to modify the switching behavior of adjacent ferromagnets (FM) through the exchange bias effect[3], or are used to tailor their interactions in dense nonvolatile information storage[4]. Antiferromagnets posses a number of interesting properties in this area such as robustness against perturbations due to magnetic fields, absence of parastatic stray fields, ultrafast dynamics (up to THz ranges)[5] and large magnetotransport effects[6]. It is thus desirable to tune and understand the spin structure and properties of nanometric antiferromagnetic films not only for applications but also to get insight into fundamental magnetic phenomena[7–10].

In this work, we focus on the antiferromagnetic insulators NiO and CoO. Their Néel temperatures are 525 K and 293 K respectively[11]. They both have the rock-salt structure with similar lattice spacings (0.416 nm and 0.426 nm respectively[12]) and present a type-II antiferromagnetic ordering: the cation magnetic moments are arranged ferromagnetically within the (111) planes and alternating planes are antiferromagnetically coupled along the $\langle 111 \rangle$ directions[12]. In the case of NiO the spins lie in the (111) planes and within each one, their direction can point along one of the three equivalent $\langle 211 \rangle$ directions, giving rise to a total of 12 different domains. CoO stands out for presenting a large 1.2% tetragonal contraction[12] below the Néel temperature. The antiferromagnetic domain distribution in CoO is more complex and still under debate[13]. Both NiO and CoO have sizeable magnetic orbital moments of 0.17 $\mu_B$ and 0.25 $\mu_B$ per formula unit respectively[14–16]. Changing the composition of a mixed Co-Ni oxide thus opens the way to tune the properties of the material between those of CoO and NiO, potentially allowing to adjust the Néel temperature between 293 and 525 K, in order to obtain a film with composition and moments close to those of CoO but with antiferromagnetic order well above room temperature.

According to the Co-Ni-O phase diagram, there is a large miscibility region where a mixed Co-Ni rocksalt phase should be stable[17]. In fact, Carey and Berkowitz[18] used $Co_x Ni_{1-x}O$ films to modify the exchange coupling with ferromagnetic layers, suggesting that the Néel temperature could be modified in this way.

Here we report on the growth and characterization of $Ni_x Co_{1-x}O$ (NCO) nanostructures prepared by high temperature oxygen-assisted molecular beam epitaxy (HOMBE) on Ru(0001). Using polarization-dependent x-ray absorption spectromicroscopy at the Co and Ni $L_{3,2}$ edges ($2p \rightarrow 3d$ transitions), we observe strong antiferromagnetic contrast at room temperature (RT). Temperature-dependent measurements identify the magnetic contribution to the observed contrast in linear dichroism and allow to determine the Néel temperature of samples with different composition.

[1]Instituto de Química Física Rocasolano (CSIC), Madrid, E-28006, Spain. [2]Alba Synchrotron Light Facility, CELLS, Barcelona, E-08290, Spain. [3]Dpto. de Física Aplicada, Universidad Autónoma de Madrid, Madrid, E-28049, Spain. [4]Instituto de Cerámica y Vidrio (CSIC), Madrid, E-28049, Spain. Correspondence and requests for materials should be addressed to A.M. (email: ania.mandziak@iqfr.csic.es)

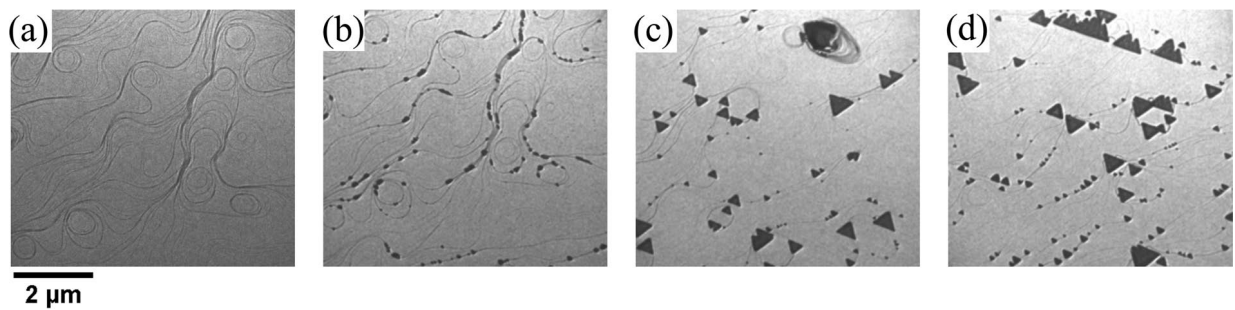

**Figure 1.** Selected frames from a sequence of LEEM images acquired during the growth of $Ni_xCo_{1-x}O$ (ratio Ni:Co 1:2) at 1150 K at a background pressure of $1 \times 10^{-6}$ mbar of molecular oxygen. The frames correspond to (**a**) 0 sec, (**b**) 3 min, (**c**) 20 min and (**d**) 45 min after opening the dosers shutter. The electron energy is 5 eV.

## Results and Discussion

Nickel oxide or cobalt oxide films have previously been grown by oxygen-assisted molecular beam epitaxy on different oxide substrates such as MgO[19–25]. Less common is the growth on metallic substrates where Ir(100)[26] or PtFe[27] have been used, in addition to Ru(0001)[28,29]. Both oxides have also been grown in order to study the proximity effect at their common interface[25,30]. However, the growth of thin films of mixed Ni-Co oxides, to the best of our knowledge, has been only attempted when trying to grow nickel cobaltite[31–33].

In the following, we first discuss our observations of the growth of mixed nickel-cobalt oxides on Ru(0001) at high temperatures focusing on the sample morphology and islands shape using LEEM. Next, we study the chemical and magnetic properties of different microstructures by x-ray absorption spectromicroscopy. Finally, we focus on the composition dependence of the Néel temperature.

**Growth mode and particle shape.** We deposit nickel and cobalt on Ru(0001) at elevated temperature (1150 K) while exposing it to molecular oxygen. As a first step the bare Ru substrate is exposed to molecular oxygen. Oxygen adsorbs on the Ru crystal at high temperature forming a 2-dimensional atomic oxygen gas, as reported previously by Piercy[34]. A sequence of images of the growth process is shown in Fig. 1. The LEEM image of the substrate just after opening the Co and Ni evaporators is presented in Fig. 1b. Many small nuclei decorate the step edges of the substrate. The dosing rate was adjusted to obtain samples with different compositions (at rates of order of $10^{-3}$ $ML_{Ru}/s$). Three samples with Ni:Co ratios of 1:2, 1:6 and 0:1 (i.e., pure cobalt oxide) were prepared. We have previously reported that the growth of CoO on Ru(0001) takes place in the Volmer-Weber mode, where 3-dimensional islands nucleate from the beginning[28]. The same growth mode is observed for the mixed NCO samples. We have not detected any influence of the Ni:Co ratio on the growth mode. As in the case of pure CoO islands, mixed oxide islands mostly present a triangular shape with two opposite orientations[28]. Similar triangular islands are also observed when growing spinels ferrites containing $Fe^{3+}$ cations on Ru (0001) by the same method[35]. Additionally, some rectangular shaped islands are found.

We first study the surface structure of each island type by LEED. The diffraction patterns are presented in Fig. 2b–d. The brightest spots correspond to the 1$^{st}$ order Ru diffracted beams. Adsorbed surface oxygen gives rise to a $2 \times 2$ LEED pattern at room temperature (marked with a magenta diamond). In addition to the patterns arising from the bare ruthenium and the oxygen superstructure, we can distinguish a hexagonal pattern (marked with a green hexagon), rotated clockwise by 5° with respect to the Ru one and a square one (marked with a blue square).

We attribute the hexagonal pattern to the triangular islands, and the square pattern to the rectangular ones. Performing micro-spot LEED by using an aperture to select the area from which the pattern is obtained (areas for $\mu$-LEED are shown as insets in Fig. 2c,d), we confirm this assignment. The two kind of islands thus correspond to different crystallographic orientations. On the basis of LEED patterns we can estimate the lattice spacing for each island type. The lattice spacing for the triangular islands is $(0.29 \pm 0.05)$ nm, slightly larger than the Ru(0001) value, 0.27 nm (see Fig. 2c). The in-plane lattice spacing for the (111)-oriented rock-salt phases should be 0.29 nm for NiO and 0.30 nm for CoO, in good agreement with the observed value. Thus, we identify the triangular islands as islands oriented along the (111) plane. The square pattern corresponds to a lattice spacing of $(0.31 \pm 0.05)$ nm, to be compared with the expected value for CoO(001) of 0.32 nm. This suggests that the square-shaped islands have the (001)-orientation. It has to be noted that not all the possible domains are observed in the LEED patterns of Fig. 2b, because of the reduced number of islands in the field of view.

Besides for CoO and $Ni_{1-x}Co_xO$, also for cerium oxide on Ru(0001) it was reported[36] that islands with two orientations, (001) and (111), can be grown depending on the oxygen background pressure. In an oxygen-rich environment the ceria islands tend to be grown in the $CeO_2$ (001) orientation, while an oxygen lean-environment favors the (111) one[37].

The thickness of the islands can be estimated from the shadows observed in the XPEEM images (see Fig. 2e). Such images are acquired by collecting the low-energy secondary electrons emitted by the sample upon irradiation with x-rays. As x-rays are hitting the sample at a polar angle of 16° (with respect to the surface plane), thick enough islands will cast a measurable shadow on the substrate. Typical heights for the observed islands are tens of nanometers or less (e.g. the island shown in Fig. 2e is 70 nm high).

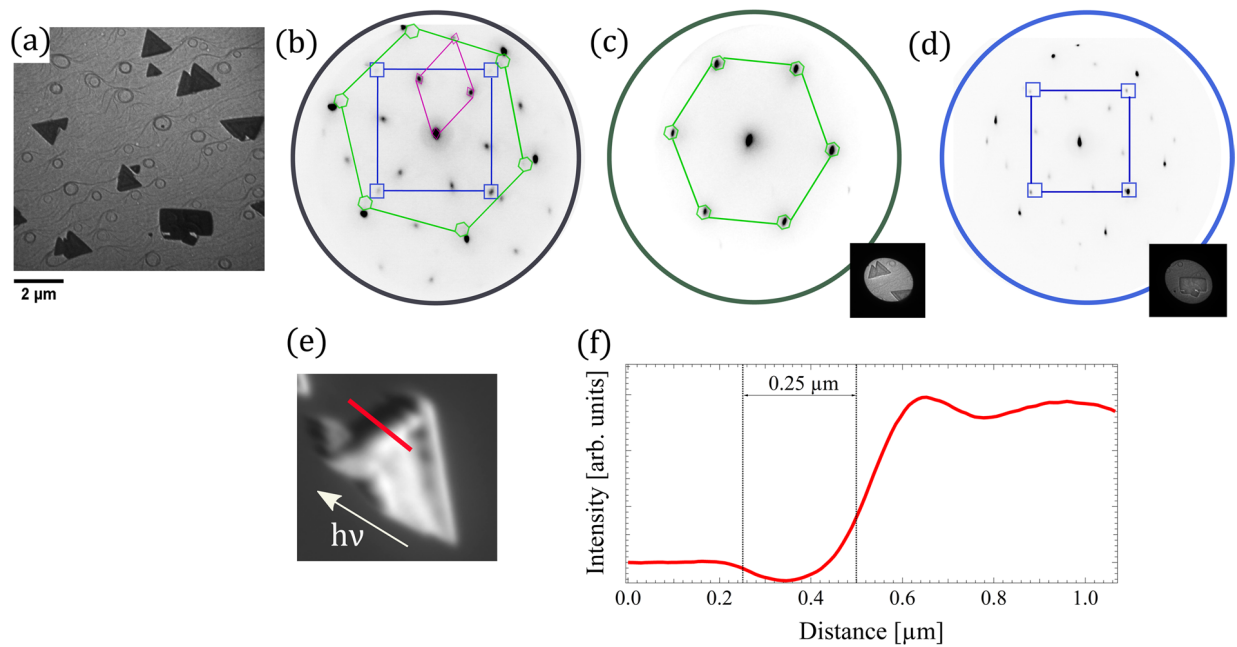

**Figure 2.** LEEM image (**a**) and corresponding LEED patterns at 45 eV electron energy from different regions on the sample (ratio Ni:Co 1:2) (**b**) substrate plus islands, (**c**) triangular islands, (**d**) rectangular island. The corresponding selected areas for the LEED are shown as insets. (**e**) XAS image acquired at photon energy close to the maximum of Co $L_3$ absorption edge. (**f**) Cross section through the island edge in the XAS image shown in (**e**).

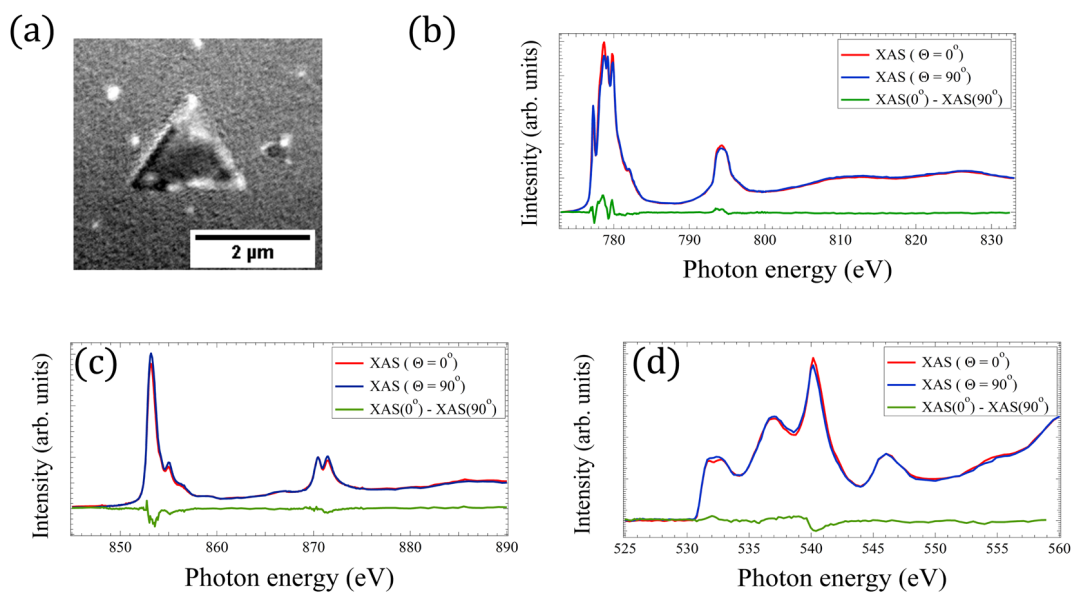

**Figure 3.** XMLD PEEM image of a $Ni_x Co_{1-x} O$ (ratio Ni:Co 1:2) island on Ru(0001) (**a**) acquired at the Co $L_3$ edge. Experimental polarization dependent (**b**) Co, (**c**) Ni and (**d**) O XAS and XMLD spectra acquired at the dark region of the triangular island presented in (**a**). $\theta$ is the angle between the light polarization vector and the surface.

**Chemical characterization by XAS-PEEM.** From stacks of PEEM images collected at different photon energies, spanning the Co and Ni $L_{3,2}$-edges the XAS spectra can be extracted from selected areas on the surface. This was done for example for the island shown in Fig. 3a. Figure 3 shows spectra at the Co $L_{3,2}$ (3b), Ni $L_{3,2}$ (3c) and O K (3d) edges XAS for linear horizontal ($\theta = 0°$) and linear vertical polarizations ($\theta = 90°$). It should be pointed out that while for $\theta = 0°$ the polarization is fully in the surface plane, for vertical polarization of the synchrotron radiation it is not strictly perpendicular, as the photons incidence angle is 16° with respect to the surface. The difference between them gives a clear dichroic signal. Linear dichroism measures the charge anisotropy associated with both the local crystal field and the local exchange field through spin-orbit coupling[38]. The latter contribution is known as X-ray Magnetic Linear Dichroism (XMLD). The XAS spectra can be used as well to obtain information on the cation oxidation state. The triangular and rectangular (not shown) islands present the same spectra. $Co^{2+}$ in CoO has a characteristic four peak structure at the $L_3$ edge[27] and $Ni^{2+}$ in NiO has a characteristic double peak structure at both edges[24]. Additionally the spectra at the oxygen K edge were acquired for both polarizations from a single domain area. The spectra together with their difference (i.e., the X-ray Linear Dichroism (XLD) signal) are shown in Fig. 3d. The oxygen spectrum is somewhat similar to that of nickel oxide reported by Arenholz[23] with a peak at a photon energy of 540.2 eV. Thus the chemical identity of different islands is confirmed to correspond to $Ni_{1-x}Co_xO$.

|  | Evaporator flux | XAS edge jump (corrected by taking into account linear absorption coefficient) (arb. units) |
|---|---|---|
| $Ni_{0.33}Co_{0.67}O$ | | |
| Ni | $8.3 \times 10^{-4}$ ML/sec | 1.15 |
| Co | $1.6 \times 10^{-3}$ ML/sec | 2.34 |
| $Ni_{0.14}Co_{0.86}O$ | | |
| Ni | $5.5 \times 10^{-4}$ ML/sec | 0.71 |
| Co | $3.3 \times 10^{-3}$ ML/sec | 4.25 |
| CoO | | |
| Co | $1.6 \times 10^{-3}$ ML/sec | 2.78 |

**Table 1.** The Ni/Co flux ratio used during growth and the estimated values obtained by analysis of the XAS data for the samples.

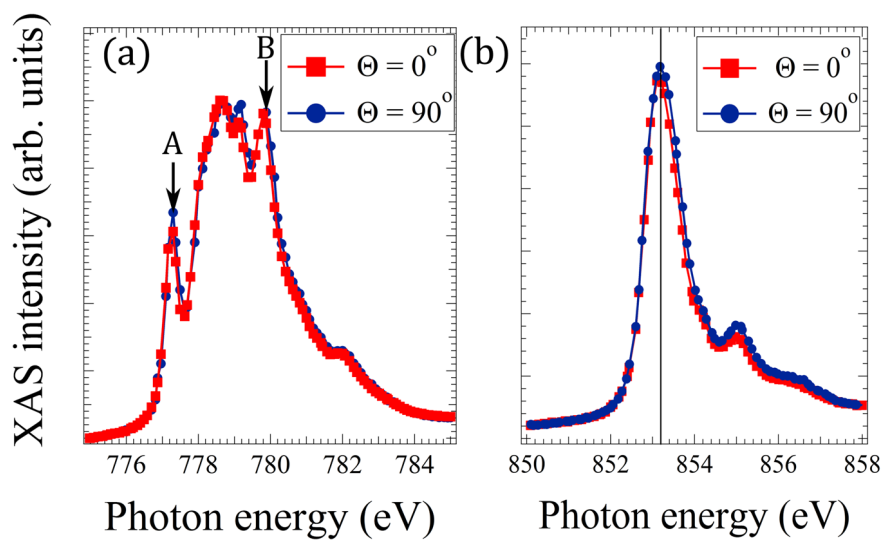

**Figure 4.** Experimental polarization dependence of the (**a**) Co $L_3$-XAS and (**b**) Ni $L_3$-XAS of a NCO (ratio Ni:Co 1:2) films on Ru(0001).

The ratio can be roughly estimated by measuring the edge jump of the XAS spectra which is proportional to the number of absorbing atoms in the sample and the linear absorption coefficient[39]. Estimated values are in good agreement with the deposited ratio for both samples (shown in Table 1). It should be noted that the probing depth is limited to the near surface region due to the short escape depth of photoelectrons (much more shorter than the x-ray penetration depth)[40,41]. Thus the composition of each sample is actually measured for the top few nm.

In order to check wether the dichroic signal in our mixed nickel-cobalt oxide samples comes from an antiferromagnetic ordering, we have studied the Ni and Co $L_3$ peak positions. It is known that if the symmetry of the material under investigation is exactly cubic, no crystal field dichroism will occur[20] and thus any observed contrast must arise from magnetic ordering. However, if the structure deviates from cubic symmetry (because of epitaxial strain in a thin film), the crystal field dichroism has to be taken into account[20,21]. In order to rule out the latter contribution we have examined the energy shift $\Delta E$ in the main peak of the Ni and Co $L_3$ white lines between $\theta = 0°$ and $\theta = 90°$. A close-up of this region is given in Fig. 4. There is no measurable shift between the two spectra acquired at horizontal and vertical polarization in the case of the Co $L_3$ peak, and for the Ni one. Thus the crystal field effect is not observable in our systems. This is already a strong hint that the observed contrast is dominated by magnetic dichroism, directly related to the presence of long-range antiferromagnetic spin order in $Ni_{1-x}Co_xO$. This is further confirmed by the different domain configuration found at the oxygen edge and by the temperature dependent measurements shown below.

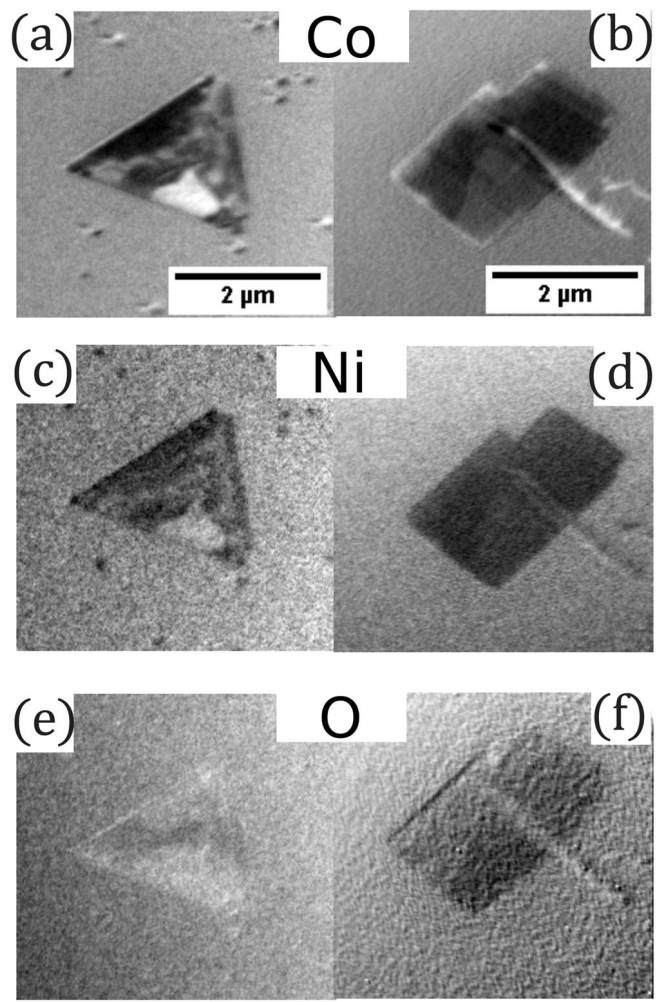

**Figure 5.** XMLD PEEM images for NCO/Ru(0001) measured at (**a**,**b**) Co $L_3$, (**c**,**d**) Ni $L_2$ and (**e**,**f**) O K edges for both types of islands found on the surface.

The situation is quite different for oxygen. In NiO, each O atom is surrounded by six Ni atoms whose moments are antiferromagnetically aligned. Thus the O XLD signal shown in Fig. 3d cannot be due to a magnetic effect. Kinoshita *et al.*[42] showed that XLD is then due to crystal distortion and can be therefore employed to map the structural domains in the near surface region.

The domain patterns shown in Fig. 5 were obtained with linearly polarized light at two energies corresponding to the maximum and minimum of XMLD signal at the Co $L_3$ edge, two energies at Ni $L_2$ edge, and at the maximum of dichroic signal at the O K edge respectively. The strongest magnetic contrast was achieved by calculating the asymmetry of two images at energies already mentioned[43] which measure the charge anisotropy associated with both the local crystal field and the local exchange field through spin-orbit coupling[27]. The distribution of the domains observed at Co and Ni edges correlate with each other. Additional to big triangular islands, small single-domain islands are also observed. However, the domains visible at the oxygen edge are different. This is a strong indication that dichroic contrast at Co and Ni edges is indeed of magnetic origin, since the cation edge dichroic images are sensitive to domains with different spin directions in addition to different ordering directions, while the anion edge is only sensitive to the latter[23,44]. The determination of the spin and ordering axis of each domain is beyond the scope of this manuscript and will be published elsewhere. Here we emphasize how the high crystalline quality of our films results in domains of up to micrometer size, much larger than typically found in NiO and CoO thin films[44].

**Néel temperature dependence on composition.** The definitive proof of the magnetic origin of the observed linear dichrosim is to heat the film above the ordering temperature of the material, i.e. the Néel temperature. In the case that the observed contrast disappears above a certain temperature, one can conclude that the contrast must be related to magnetic ordering. In the case of mixed nickel-cobalt oxides, the precise Néel temperature is not known but it is expected to be between the values for the pure oxides (293–525 K). As we already see a contrast at RT, we can expect that by adjusting the Ni:Co ratio we should be able to increase the Néel temperature above the CoO one.

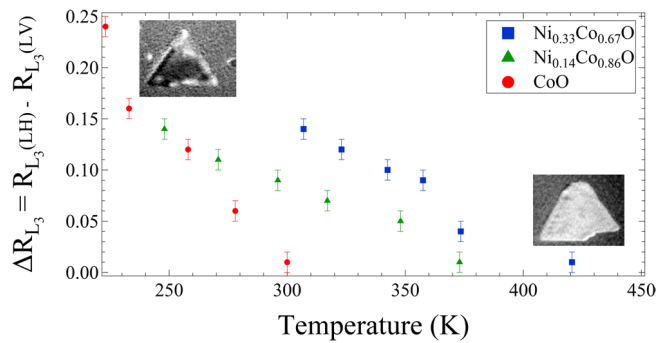

**Figure 6.** Temperature dependence of the asymmetry measured as a Co L$_3$ ratio difference $\Delta R$ L$_3$ for samples with different composition. In the inset XMLD PEEM images at the Co L$_3$ edge at RT (upper left) and above the Néel temperature (bottom right).

We now determine the Néel temperature for each sample. Following Wu et al.[45] and[25], we employ the intensity ratio $R$L$_3$ between the first and last peak at the Co L$_3$ edge (A and B, marked by arrows in the inset of Fig. 4) in order to follow the temperature dependence. It is proportional to the overall anisotropy, which tracks the magnetic order. Thus we plot $\Delta R$ L$_3$ = $R$ L$_3$(0°) − $R$ L$_3$(90°) vs temperature (Fig. 6). Using these plots, we determine the Néel temperature for each sample. For a Ni:Co ratio of 1:2 the transition temperature is 425 K, for Ni:Co 1:6 $T_N$ = 375 K and for CoO it is 295 K, very close to that of bulk CoO. We thus have shown that the Néel temperature can be modified at will by selecting the proper Co:Ni ratio.

A similar behavior was observed for epitaxial NiO/CoO films grown on MgO substrates[25]. Therein the enhancement of the Néel temperature was induced by a proximity effect. The $T_N$ of CoO in a bilayer was higher than without the NiO layer and reached a maximum at the interface owing to the interfacial nature of the exchange coupling. In our system there is a mixture of Co and Ni cations within the rocksalt structure. So now the superexchange interaction will depend on the particular cation distribution around a given one. One can expect that the exchange constant $J_{ij}$ for nearest neighbours will be between the −0.69 meV value of NiO and −0.71 meV of CoO, and between −8.66 meV and −6.30 meV for next-nearest neighbors, respectively. Based on the calculated magnetic exchange interaction constants one can conclude that by mixing two ions with different $J_{ij}$ we can tune the interaction between them and thus modify $T_N$ in a controlled way[46,47].

In order to investigate if $T_N$ depends on the film thickness, we have used the same method for islands with different thickness (between 20 and 80 nm). Our results do not show any thickness dependence, confirming previous experiments[48–50] which claimed that finite size effects appear only in films thinner than 10 nm. Note that the height of our mixed oxide islands is tens of nanometers, i.e., above such limit.

## Conclusions

To summarize, the growth by high-temperature oxygen-assisted molecular beam epitaxy of mixed cobalt-nickel oxide structures on Ru(0001) surface gives rise to the formation of mostly (111) oriented islands. The structures are easily recognized by their triangular shape and they coexist with a few (001) minority orientation rectangular islands. The atomic structure has been characterized with LEEM and μ-LEED confirming the existence of these two orientations. Chemical states of the surface cations (Co$^{2+}$ and Ni$^{2+}$) have been identified by x-ray absorption spectromicroscopy. We have shown that the growth at high temperature leads to the formation of the mixed nickel-cobalt monoxide phase as expected from the phase diagram. The element-specific XMLD measurements reveal sizeable antiferromagnetic domains at room temperature. Our results demonstrate the possibility to increase the Néel temperature of the Ni$_x$Co$_{1−x}$O antiferromagnetic nanostructures well above room-temperature and tune it by adjusting the composition, without involving any additional magnetic layer. No thickness dependence of $T_N$ was found in the range 20 to 80 nm.

## Methods

Experiments have been performed at the CIRCE-XPEEM beamline of the Alba Synchrotron Light Facility[51]. The microscope can work in a pure Low-Energy Electron Microscopy (LEEM) mode, allowing also diffraction measurements by using μ-LEED mode. By using synchrotron light it can work as an X-ray PhotoEmission Electron Microscope (XPEEM). It allows to acquire images of the energy-filtered photoelectron distribution from micro-sized selected areas of the surface, with an energy resolution below 0.2 eV. The sample is illuminated by incoming photons at an angle of 16° from the surface plane. The sample azimuth relative to the x-ray beam can be changed at will. The kinetic energy of the photoelectrons used to form an image can also be selected. In addition to X-ray Absorption Spectroscopy (XAS) images, dichroic images can be obtained by measuring the pixel-by-pixel asymmetry between images with orthogonal x-ray polarizations directions[38,52].

Mixed nickel-cobalt oxide films were deposited onto a (0001)-oriented ruthenium single crystal. The method of cleaning the Ru crystal is described elsewhere[53]. Nickel-cobalt oxide films were synthesized by oxygen-assisted MBE, i.e., co-depositing Ni and Co onto Ru at elevated temperature (1150 K) in a molecular oxygen atmosphere ($1 \times 10^{-6}$ mbar). Nickel and cobalt dosers were previously calibrated by measuring the time needed to complete a monolayer on the W(110) and Ru(0001) surfaces respectively. Typical rates used in the experiments were $1.6 \times 10^{-3}$ ML$_{Ru}$/s for Co and $8.3 \times 10^{-4}$ ML$_{Ru}$/s for Ni for a 2:1 Co:Ni ratio (where 1 ML$_{Ru}$ is defined as $1.4 \times 10^{19}$ atoms/m$^2$).

Growth was performed during *in-situ* observation by LEEM in order to optimize the growth parameters such as substrate temperature, evaporator flux and oxygen partial pressure. Large 3-dimensional islands were found at 1150 K. After the growth process, the sample was cooled down to room temperature in an oxygen atmosphere. The x-ray absorption experiments were typically performed a few days after growth. Before synchrotron measurements, the sample was flashed up to 900 K in $10^{-6}$ mbar oxygen atmosphere to remove adsorbates while avoiding a possible reduction of the film.

## Data Availability

The datasets generated during and/or analyzed during the current study are available from the corresponding author on reasonable request.

### Acknowledgements
This work is supported by the Spanish Agencia Estatal de Investigación (MCIU/AEI/FEDER, EU)) through Projects Nos MAT2015-64110-C2-1-P, MAT2015-64110-C2-2-P, RTI2018-095303-B-C51, and RTI2018-095303-B-C53, by the European Commission through Project H2020 No. 720853 (Amphibian) and by the Comunidad de Madrid through Project. NANOMAGCOST-CM P2018/NMT-4321. These experiments were performed at the CIRCE beamline of the ALBA Synchrotron Light Facility. A.M. acknowledges funding via a CSIC-Alba agreement.

### Author Contributions
J.d.l.F. and L.A. conceived the experiment, A.M., G.D.S., C.G.-M., A.Q., P.P., J.E.P., J.d.l.F., L.A. and M.F. conducted the experiments. A.M. and J.d.l.F. analyzed the results. A.M. lead the manuscript writing with contributions from all the authors. All authors reviewed the manuscript.

### Additional Information
**Competing Interests:** The authors declare no competing interests.

**Publisher's note** Springer Nature remains neutral with regard to jurisdictional claims in published maps and institutional affiliations.